# The Need for Further Development of Magnetrons as RF Sources for HEP

Thomas Kroc, Vyacheslav Yakovlev, Charles Thangaraj, Brian Chase, Ram Dhuley

High Energy Physics complexes are large users of energy. The HEP community has long been cognizant of this energy usage. In the early development of the Fermilab Tevatron, it was recognized that superconducting magnets could provide a reduction in energy usage of the Main Ring and Energy Saver from 65 MW to 17 MW, depending on operating parameters [1]. This awareness continues as evidenced by the recent PIP-II Sustainability Workshop [2]. In this light, we should continue to exploit all possible developments that can increase the efficiency of the energy that we use.

Other than the energy used to power and cool magnet systems, the RF power used to accelerate beams is a very large consumer of energy. While solid state RF power sources offer many advantages, they are inherently poor performers with regard to efficiency, with maximum values typically less than 60% in terms of wall plug to RF power.

Magnetrons are inherently high efficiency devices. Recent advances allow them to be used as RF sources for accelerating cavities. We propose that the HEP community embark on a development program to complete this work in order to capitalize on the many benefits that magnetrons can provide in efficiency, size, and power. Since the magnetron's efficiency is almost constant from 30% to 100% of the design power, the power can be tuned close to that required for a particular accelerator cavity, with phase modulation producing fast corrections in the amplitude. This differs from a klystron, where operation at significantly less than the optimum power results in decreased efficiency.

A thorough analysis of the efficiency of accelerator systems has been performed [3]. As an example, this analysis has been updated in Figure 1 to illustrate the gains that could be achieved in the PIP-II complex by the use of magnetron RF sources. As shown by the arrows, present solid-state RF supplies are approximately 45% efficient, compared to 85% for magnetrons. However, magnetron sources suitable for HEP are not yet ready. While phase and amplitude locking [4] now makes it possible to link the free running, self-excited magnetrons to RF cavities, a number of parameters, such as magnetron performance, unit-to-unit central frequency, mean lifetime, and various supporting power supplies need to be improved to meet HEP reliability needs.

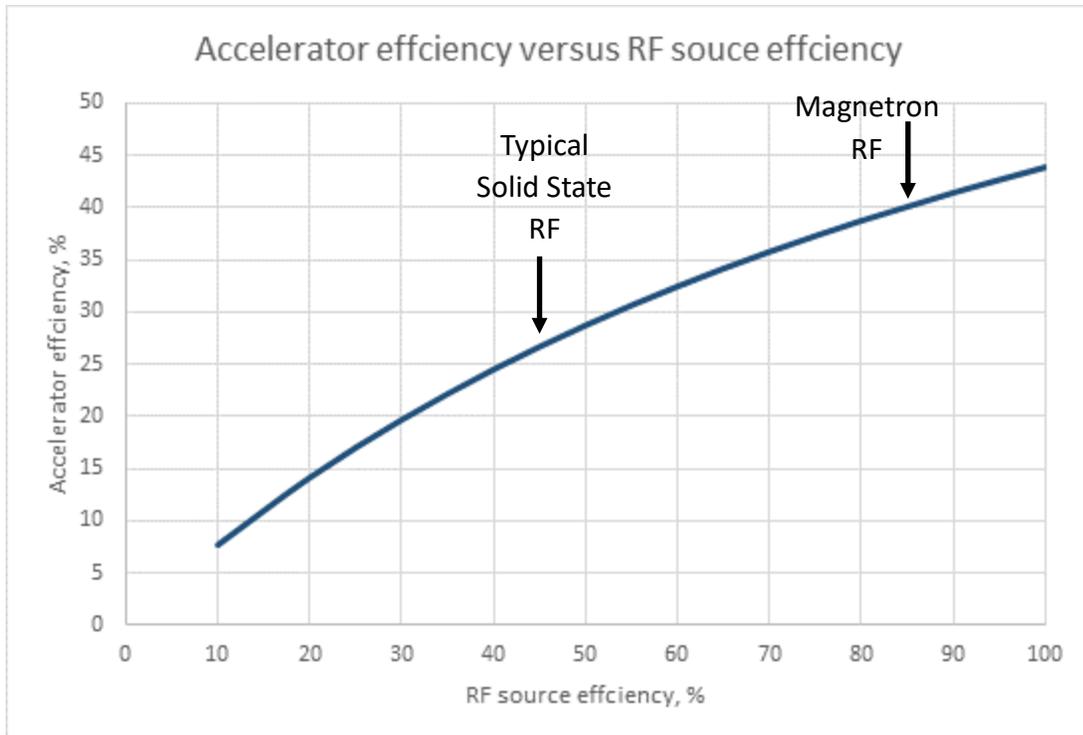

*Figure 1 Using PIP-II as an example, comparison of accelerator system efficiency with solid-state and magnetron RF sources.*

Improving the performance of high-power magnetrons will allow us to be responsible environmental stewards in our pursuit of discovery science. A frequently cited benefit of basic science is spin-offs of technology. These improved magnetron sources will then be available as efficient sources for many industrial accelerator systems that presently rely on less efficient sources. For instance, two recent design studies of industrial high-power SRF e-beam industrial accelerators have shown their cost and energy efficiency to be limited by the RF source [5,6]. These would greatly benefit from the availability of high-efficiency magnetron RF sources.

Solid State Amplifiers

Historically, vacuum electron tubes have been used for medium and high-power applications with solid state amplifiers used as driver stages. These driver stages had limitations on the maximum frequency at which they could operate. Developments in MOSFET and CMOS technology now covers all frequencies presently of interest to HEP (Figure 2).

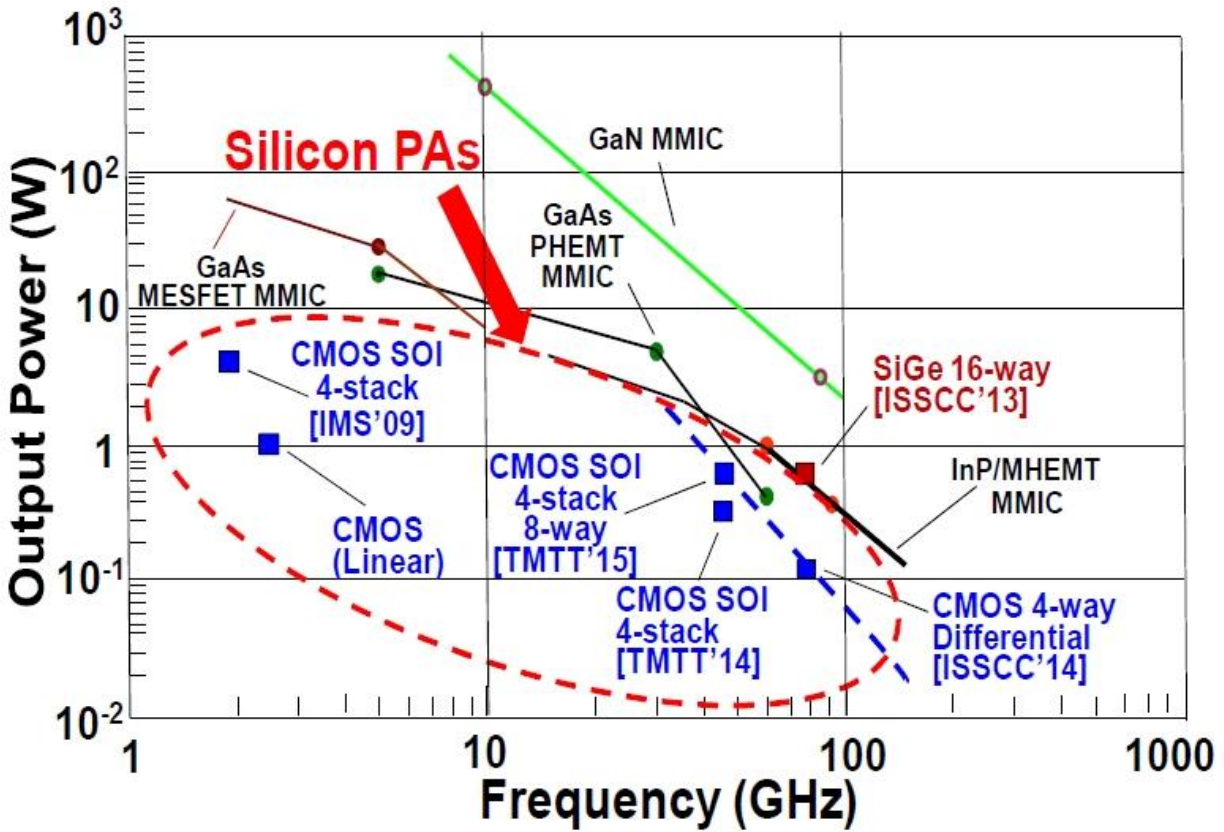

*Figure 2 Power range and frequency capabilities of present solid-state RF sources.*

Vacuum tube technology can produce individual components with peak powers in the 5 – 12 MW range while supporting duty factors of a few percent. The output of these amplifiers can be fed directly into large accelerating cavities such as drift tube or side-coupled linacs.

Solid state power amplifiers rely on a large number of low power modules whose output must be combined to achieve high power. The individual solid-state modules have efficiencies of approximately 55%. The need for a large number of combiners or circulators further reduces the efficiency of the system. Overall system efficiencies therefore are less than 50%. This large number of contributing components results in large devices as seen in Figure 3 of a 2 MW solid state system. These solid-state systems are also expensive at over $10/Watt.

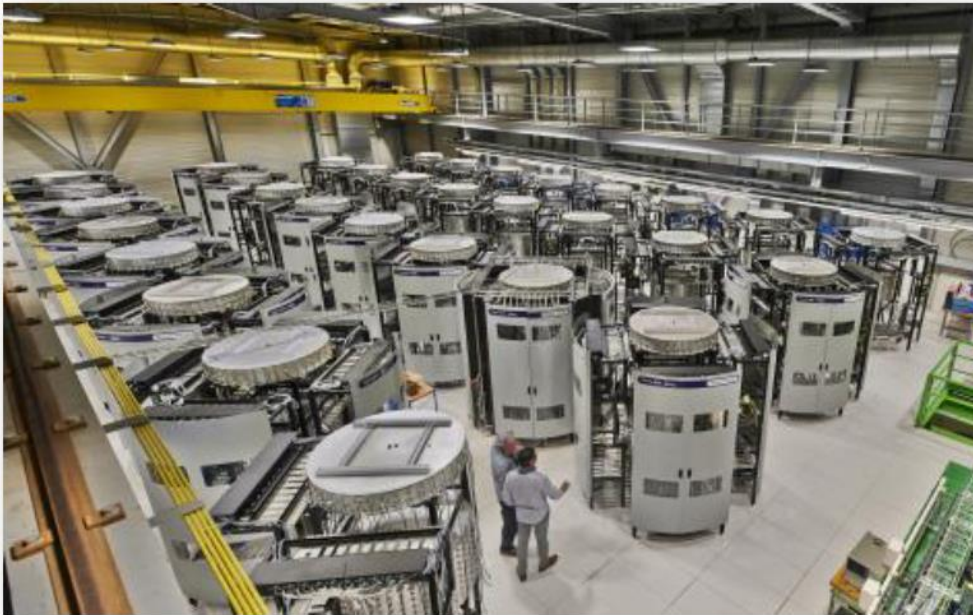

*Figure 3* CERN SPS Thales Solid State Power Amplifiers, 32 towers x 144 kW @ 200 MHz, combined into 2 x 2 MW amplifiers, into operation since 2021

Magnetrons

High-power magnetrons are commercially available and are used by the process heating industry. These devices offer power of 100 kW at 915 MHz. They are inexpensive at less than $10,000 each or $0.10/Watt. However, reported lifetimes are 6,000 hours or less with some degree of variability. Also, the unit-to-unit variability of the central frequency can vary by a few MHz. This performance is not suitable for HEP needs.

The low cost and high efficiency provide great motivation to improve these parameters for HEP. The cost advantage is so great that if the improvements noted below significantly increase their cost, there will still be a great savings over solid state.

Significant work has been conducted that allows amplitude and phase control of magnetrons allowing them to be used as RF sources for accelerating cavities. [4,7-9] This work began utilizing a transmitter consisting of two 2-cascade injection-locked magnetrons whose outputs were combined in a 3-dB hybrid. The wideband dynamic management of the output power of the transmitter model was first experimentally demonstrated using 'combined in power' magnetrons, injection-locked by the phase-modulated signals. Experiments with the injection-locked magnetrons adequately emulated the wideband dynamic control with a feedback control system, which would allow the suppression of all known parasitic modulation of the accelerating field in the SRF cavities.

This concept was experimentally demonstrated using a 100 kW, 1300 MHz magnetron. [4] The system produced 100 kW at 1.3 GHz with 1.5-ms pulses. The duty factor was limited only by the availability of ancillary power supplies.

Between 2012 and 2021 Fermilab APS-TD in collaboration with Muons, Inc. has conducted additional research to apply magnetrons to SRF linacs. This work has had the following results:

- Development and testing of:
    - A cascade scheme to improve the gain of a magnetron-based RF system.
    - A para-phasing amplitude and phase modulation scheme.
    - magnetron operation below critical voltage at high locking signal.
- Experimental work has been conducted that allowed understanding mechanisms of the magnetron operation in injection-locked regime below critical voltage.
- Based on this, it has been shown experimentally that it is possible to operate a magnetron below critical voltage with a high locking signal, which in turn, gives the following possibilities:
    - Operation of the magnetron at different output power with very high efficiency, >80%. It is essential for proton/ion SRF linacs, where the cavity input power varies along the linac;
    - Magnetron operation in this regime provides very narrow spectrum of the output signal, which is very important for SRF linac application.
    - Precise phase and amplitude control of the output power, which is essential for the linac, especially if the accelerated beam is injected to the ring for accumulation or further acceleration.
    - It is shown experimentally that operation below critical energy at high locking system provides higher efficiency.
    - Moreover, there are experimental evidence of a magnetron longevity improvement in this operation regime.
- Different mechanisms of the considerably short magnetron life span have been analyzed and the ways to improve longevity has been suggested
    - external pumping,
    - RF circuit improvements.
- Different schemes to get high locking signal have been suggested:
    - cascade scheme for multi-cavity accelerators (tested)
    - reflector for single-cavity industrial accelerators (to be tested).

These results confirm that the magnetron is an excellent candidate as an RF source for SRF linacs due to their low cost and simplicity.

It should be noted that recent funding announcements for the DOE's Accelerator Stewardship Program have included in its topics a call for High Efficiency High Average Power RF sources with the goal of devices with efficiency greater than 80% and average power of at least 250 kW. The funding available, and the goals of the Stewardship Program (with regard to target TRL levels) are mainly suitable for initial design work and specifically, cost estimates. However, to be fully realized, a complete roadmap needs to be developed, including target frequency choices and power milestones.

Magnetron Development Program

As noted, a major issue that inhibits the use of magnetrons is their low, and variable life span. Magnetron tubes are inexpensive, so the cost of replacement is not a concern for the process heating industry (drying wallboard and lumber) which is a major user of magnetrons. For HEP, the concern would be process interruption on a frequent or irregular basis. Longer lifetime and a more dependable mean-time-to-failure is necessary for adoption of this power source. The main reasons for the present state of tube lifetime are:

- Anode sputtering of the cathode material.
- Cathode bombardment by backward electrons.

General measures which may be taken to address these issues are:

- Active vacuum pumping of the magnetron.
- Electron dynamics optimization.

Attention to electron dynamics may also improve the magnetron efficiency. Recent investigations show [10, 11] that magnetron efficiency may be improved together with lifetime extension by operating in a sub-critical operation regime with a larger locking signal.

IARC at Fermilab's function is to transfer technology developed for the science projects, by Fermilab and DOE overall, to industrial applications. An important project for IARC is the development of a compact, super conducting RF (c-SRF) electron accelerator. Fermilab's strength in this area is its expertise in conduction cooled, superconducting, $Nb_3Sn$ coated cavities. IARC's current efforts are in two areas (i) integrating conduction cooled superconducting cavities with commercially available technology and (ii) integrating conduction cooled superconducting cavities with new technology developments (i.e., integrated electron guns and a low heat loss coupler). For these efforts we are using solid state RF sources, as, we believe that this is the best choice at this time. But we also believe that for the long term, five or ten years from now, magnetrons may be key for efficient, disruptive electron accelerators for industrial use. With these thoughts in mind, we have outlined a program to address these issues of magnetron lifetime and efficiency.

The program:

1. Use the 3D simulation code, MICHELLE [12], to understand in detail the beam dynamics of a magnetron. This will allow self-consistent beam modeling of the electron flow in a magnetron in 3D RF and DC magnetic field in presence of the space-charge limited current emission. The emission model will need to be modified to take into account tangential DC magnetic fields on the cathode. These modifications to the emission model have been developed and tested for high-power electron guns for electron cooling [13] and should be implemented in MICHELLE. The code should be modified also to simulate the transient processes of tube excitation and operation regimes – like it has been done for IOT modeling [14]. This effort would involve a partnership with Leidos (former SAIC), the developer of the MICHELLE code.

2. Next, the code will need to be benchmarked with the collaboration of a company with experience in high-power magnetron development. The partner would need to provide drawings of an existing well-measured magnetron, which would be simulated using the modified MICHELLE code. This improved and benchmarked code will strengthen the national RF industry allowing better designs of the magnetron for different applications – scientific, industrial, civil, and military.

3. Finally, it would be possible to optimize the magnetron design to improve its longevity and efficiency and optimize various operation regimes. Different options could be explored, like 2D harmonic cavities, different types of cathodes including the newly developed Nanocomposite Scandate Tungsten cathodes [15].

The goal would be to achieve an efficiency of more than 85% with tube lifetime of ~50,000-80,000 hours.

System Development

In addition to the development of better performing magnetrons themselves, further work on complete RF systems is needed. As noted in Reference 4, a 100-kW system has been assembled and operated at full power but at a low duty factor. Appropriate funding and focus is needed to better characterize this or similar systems. Other elements of the RF system also need development to support magnetron operation. For instance, DC power supplies are a large part of the ancillary components but have stringent regulation requirements.

Summary

High Energy Physics will always continue to grow in the energy and intensity of its accelerator systems. The RF power required will grow proportionally. While a number of facilities are turning to solid state RF power sources, the size and cost of these systems are burdensome. The inefficiency of solid-state sources should make them unacceptable in light of growing climate concerns.

Magnetron RF sources can provide highly efficient, compact, high-power devices. By providing large amounts of power per device, they reduce losses due to combiners and circulators. A development program to further develop devices to match the performance parameters needed for HEP will have great benefit in the long run.

The Snowmass community can create more reliable and more energy efficient accelerators. We believe we should embark on this endeavor now, before we are requested to do so by the circumstances. A goal of more reliable accelerators for science and industry is within Snowmass community goals, mission, and capabilities.